\documentclass[superscriptaddress,twocolumn,showpacs,preprintnumbers,amsmath,amssymb,pra,10pt,aps]{revtex4-1}
\usepackage{graphicx}
\usepackage{dcolumn}
\usepackage{bm}
\usepackage{subfigure}
\usepackage{multirow}
\usepackage{soul}
\usepackage[normalem]{ulem}
\usepackage{xcolor}
\usepackage{kantlipsum}
\usepackage{textcomp}

\definecolor{amber}{rgb}{1,0.49,0}

\newcommand{\editor}[2]{%
  \expandafter\newcommand\csname #1note\endcsname[1]{%
    \textcolor{#2}{(\textbf{#1:} ##1)}}%
  \expandafter\newcommand\csname #1\endcsname[1]{%
    \textcolor{#2}{##1}}%
  \expandafter\newcommand\csname #1cancel\endcsname[1]{%
    \textcolor{#2}{\sout{##1}}}%
  \expandafter\newcommand\csname #1change\endcsname[2]{%
    \textcolor{#2}{\sout{##1} ##2}}%
  \newenvironment{#1text}{\color{#2}}{\color{black}}
}

\def\smallE{{\scriptscriptstyle E}}

\def\smallZ{{\scriptscriptstyle Z}}
\def\smallXC{{\scriptscriptstyle XC}}
\def\smallDFT{{\scriptscriptstyle DFT}}
\def\smallGGA{{\scriptscriptstyle GGA}}
\def\smallH{{\scriptscriptstyle H}}
\def\smallKS{{\scriptscriptstyle KS}}
\def\eqref#1{(\ref{#1})}

\def\rOmega{{\mathrm\Omega}}

\editor{SB}{amber}
\editor{FG}{blue}
\editor{CAVEAT}{red}
\editor{resub}{cyan}

\usepackage{mathrsfs}
\usepackage{placeins}

\def\upt{{\mathcal{T}}}
\newcommand{\stoc}[1]{\mathscr{#1}}
\def\cycle{{\tau}}

\begin{document}

\title {Invariance principles in the theory and computation of transport coefficients}
\author{Federico Grasselli}
\affiliation{
  COSMO -- Laboratory of Computational Science and Modelling, IMX, \'Ecole Polytechnique F\'ed\'erale de Lausanne, 1015 Lausanne, Switzerland
}

\author{Stefano Baroni}
\affiliation{
   SISSA -- Scuola Internazionale Superiore di Studi Avanzati, 34136 Trieste, Italy, EU
}
\affiliation{
  CNR-IOM DEMOCRITOS Simulation Center, SISSA, 34136 Trieste, Italy, EU
}

\date{\today}


\begin{abstract}
In this work we elaborate on two recently discovered invariance principles, according to which transport coefficients are, to a large extent, independent of the microscopic definition of the densities and currents of the conserved quantities being transported (energy, momentum, mass, charge). The first such principle, \emph{gauge invariance}, allows one to define a quantum adiabatic energy current from density-functional theory, from which the heat conductivity can be uniquely defined and computed using equilibrium \emph{ab initio} molecular dynamics. When combined with a novel topological definition of \emph{atomic oxidation states}, gauge invariance also sheds new light onto the mechanisms of charge transport in ionic conductors. The second principle, \emph{convective invariance}, allows one to extend the analysis to multi-component systems. These invariance principles can be combined with new spectral analysis methods for the current time series to be fed into the Green-Kubo formula to obtain accurate estimates of transport coefficients from relatively short molecular dynamics simulations.
\end{abstract}


\maketitle

\section{Introduction}
\label{sec:intro}

Transport coefficients are archetypal examples of off-equilibrium properties, which describe in fact entropy production and the approach to equilibrium in extended systems, thus giving a quantitative meaning and a conceptual framework to the very notion of irreversibility and the arrow of time. While non-equilibrium statistical mechanics is still a very active and largely unsettled field of research, the relaxation of small off-equilibrium fluctuations and the response of systems to small perturbations have been {given} a rigorous theoretical foundation back in the fifties by the  Green-Kubo (GK) theory of linear response \cite{Green1952,*Green1954,Kubo1957a,*Kubo1957b}. Among other achievements, this theory provides a rigorous and elegant way to cast the computation of transport coefficients into the evaluation of equilibrium time correlation functions of suitably defined fluxes, thus making it accessible to equilibrium molecular dynamics (EMD) simulations. This feat notwithstanding, the several conceptual subtleties underlying the linear-response theory of transport coefficients are often dodged or disguised with a clumsy notation that makes it difficult to fully appreciate their scope and impact on the design and use of computer simulation methodologies. Also due to this predicament, a number of misconceptions have affected the otherwise mature and fecund field of computer simulation of transport in condensed matter, thus unduly limiting its scope. Foremost among these misconceptions is that the intrinsic indeterminacy of any local representations of an extensive quantity, such as the density or atomic break-up of the energy (or charge and mass, for that matter), would undermine the uniqueness of the transport coefficients that are derived from them. While similar reservations should apply to classical and quantum \emph{ab initio} simulations alike, they have in fact impacted mainly the latter, to the extent that until recently it had been believed that the GK theory of heat transport could not be combined with quantum simulation methods based on electronic-structure theory. This quandary has been recently overcome, mainly thanks to the introduction of a so-called \emph{gauge invariance} principle of transport coefficients \cite{Marcolongo2016,Ercole2016}, which basically states that, under well defined conditions, the value of a transport coefficient is largely independent of the detailed form of the conserved densities and fluxes from which they are derived through the GK formulae. In full generality, gauge invariance implies that the value of a transport coefficients is unchanged if the flux from which it is calculated is modified by adding to it a vector process whose power spectrum vanishes at $\omega=0$ (such a process is conveniently dubbed \emph{non-diffusive}). Further difficulties may arise in the case of multi-component systems, where the interaction among different fluxes make the transport of one conserved quantity (such as, \emph{e.g.}, energy), depend on the dynamics {of} the other hydrodynamical variables (such as, \emph{e.g.}, the number of different molecular species), thus muddling the definition of heat conductivity in these systems. This difficulty is solved by defining, \emph{e.g.}, the thermal conductivity as the ratio between the energy flux and the temperature gradient, when all the other conserved fluxes vanish. In this case, a further invariance principle, dubbed \emph{convective invariance} \cite{Bertossa2019}, states that the thermal conductivity results to be unchanged if the definition of the energy flux is altered by adding to it an arbitrary linear combination of the mass fluxes of the molecular species constituting the system. While the value of the transport coefficients enjoys the invariance properties mentioned above, the statistical properties of the flux time series from which they are derived do depend on the microscopic representation of the conserved densities and current densities, thus substantially affecting the statistical error of the transport coefficient being computed. This dependence opens the way to optimizing this representation, so as to minimize the resulting statistical errors. This freedom can be exploited in conjunction with the recently introduced \emph{cepstral analysis} of the current spectra \cite{Ercole2017} to substantially reduce the length of the EMD simulations needed to evaluate a transport coefficient to a given target accuracy.

In this paper we review the concepts of \emph{gauge} and \emph{convective invariance} in the classical theory of transport in condensed matter, with emphasis on their application to the computation of transport coefficients from the GK theory of linear response and EMD, as well as some concepts and tools for the spectral analysis of the current time series, which can be used in conjunction with these invariance principles to substantially reduce the statistical errors affecting the EMD estimate of various conductivities. In Sec. \ref{sec:theory} we briefly lay down the GK linear-response theory of transport, aiming at establishing some general terminology and notations. In Sec. \ref{sec:gauge} we introduce the concept of gauge invariance, whereas convective invariance is discussed in Sec. \ref{sec:convective}. In Sec. \ref{sec:cepstral} we discuss a newly introduced spectral method (dubbed \emph{cepstral analysis}) to evaluate and systematically reduce the statistical error affecting the estimate of transport coefficients from EMD. In Sec. \ref{sec:heat} we specialize our discussion to \emph{ab initio} heat transport and report a general expression for the microscopic heat flux suitable to density functional theory. In Sec. \ref{sec:charge} we show how gauge invariance can be combined with concepts from topology to reveal some unexpected features of charge transport in ionic conductors. Sec \ref{sec:conclusions} finally contains our conclusions.

\section{Theory} \label{sec:theory}

Transport theory is essentially a dynamical theory of \emph{hydrodynamical variables}, \emph{i.e.} of the long-wavelength components of the densities of conserved, extensive, variables, which in short we refer to as \emph{conserved charges} and \emph{densities}. In a simple fluid, for instance, the conserved charges are the energy, the three components of total momentum, and the total number of molecules of each chemical species. The corresponding transport coefficients are the thermal conductivity, the viscosity, and the various diffusivities. Let $\hat Q$ be one such conserved charge, and $\hat q(\bm{r})$ the corresponding density: $\hat Q = \int \hat q(\bm{r}) d\bm{r}$. Here and in the following we adopt the convention that a hat, as in $\hat A$, indicates an implicit dependence on the system's phase-space variables, $\Gamma$: $\hat A = A(\Gamma)$, where $\Gamma=\{x,p\}$ is the set of all the atomic coordinates and momenta. When neither a hat nor the phase-space argument are present, $A$ indicates the expectation of $\hat A$ over some suitably defined phase-space distribution, $\rho(\Gamma)$ (most frequently the canonical or micro-canonical equilibrium distributions): $A = \langle \hat{A} \rangle \doteq \int A(\Gamma) \rho(\Gamma) d\Gamma$. When it will be necessary to distinguish between the expectation of a quantity, $A$, and its phase-space representation, $A(\Gamma)$, the latter will be referred to as a \emph{phase-space sample} of the former. Sometimes, we will need to indicate explicitly the implicit time dependence of a phase-space variable (no pun intended). When this is done, we will mean the phase-space variable $\hat A(t)=A(\Gamma,t)\doteq A(\Gamma_t)$, as a function of time and of the initial condition, $\Gamma=\Gamma_0$, of a phase-space trajectory, $\Gamma_t$, as determined by Hamilton's equations of motion.

Locality implies that for any conserved density, $q(\bm{r})$, a (conserved) current density, $\bm{\jmath}(\bm{r})$, can be defined, such that the two of them satisfy the continuity equation, $ \dot{q}(\bm{r},t) +~\nabla~\cdot~\bm{\jmath}(\bm{r},t) = 0$, where $\dot q$ indicates the time derivative of $q$. A current density, $\bm{\jmath}(\bm{r},t)$, satisfying the continuity equation with a conserved density, $q(\bm{r},t)$, will be said to be \emph{conjugate} to $q$. Strictly speaking, the continuity equation holds for the expectations of conserved densities and current densities, as well as for their phase-space samples, $\hat q(\bm{r},t) = q(\bm{r}, \Gamma_t)$ $\hat{\bm{\jmath}} (\bm{r},t) = \bm{\jmath}(\bm{r},\Gamma_t)$. For the sake of unburdening the notation as much as possible, we will sometimes overlook the distinction between phase-space samples and their expectations.

The Fourier transform of the continuity equation reads:
\begin{align}
  \dot{\tilde{q}}(\bm{k},t)+i\bm{k}\cdot \tilde{\bm{\jmath}}(\bm{k},t)=0, \label{eq:continuity}
\end{align}
where $\tilde{q}(\bm{k})=\int q(\bm{r})\mathrm{e}^{i\bm{k}\cdot\bm{r}} d\bm{r}$ is the Fourier transform of $q$, and similarly for $\bm{\jmath}(\bm{r})$ and any other function of $\bm{r}$. Eq.~\eqref{eq:continuity} shows that the smaller the wave-vector, $|\bm{k}|$, \emph{i.e.} the longer the wavelength, the slower the dynamics of conserved densities and fluxes. This means that, at sufficiently long wavelength, conserved densities and current densities are adiabatically decoupled from the (zillions of) other atomically fast degrees of freedom. Also, translational invariance implies that conserved densities at different wavevectors do not interact with each other. As a consequence, the dynamics of hydrodynamic variables is determined by a handful of equations that couple them with each other at fixed wavevector. When the intensive variables conjugate to the conserved quantities depend on position sufficiently slowly, the system can be thought of as locally in thermal equilibrium and conserved currents can then be connected to the \emph{thermodynamical affinities} (\emph{i.e.} to the gradients of the intensive variables) through the so-called \emph{constitutive equations} \cite{Kadanoff1963}. By combining the constitutive equations with the continuity equations of all the conserved densities and currents, the Navier-Stokes equation of classical hydrodynamics can be derived \cite{Kadanoff1963}.

Let us consider a system described by a Hamiltonian $\hat{H}^\circ$ and subject to a time-dependent external perturbation, $\hat{V}(t)=\sum_i\int v_i(\bm{r},t) \hat{q}_i(\bm{r})d\bm{r}$, where $\{\hat{q}_i\}$ is a set of conserved densities. The perturbed Hamiltonian reads:
\begin{align}
  \hat H = \hat{H}^\circ + \sum_i\int v_i(\bm{r},t) \hat{q}_i(\bm{r})d\bm{r}.
\end{align}
Here the $v_i$s are to be treated as strengths of the perturbation in linear-response theory. As such, they are \textit{not} phase-space functions, and they dependence on time only explicitly. To first order in the $v$'s, the expected value of the conserved current densities conjugate to the $q$'s, $\{\hat{\bm{\jmath}}_i\}$, can be obtained from the GK theory of linear response \cite{Kubo1957a,*Kubo1957b} as \cite{baroni2018}:
\begin{multline}
  \jmath_{i\alpha}(\bm{r},t) = -\frac{1}{k_BT} \\ \times \sum_j\int d\bm{r}'\int_{-\infty}^t dt' \Bigl\langle \hat{\jmath}_{i\alpha}(\bm{r},t) \dot{\hat{q}}_j(\bm{r}',t') \Bigr\rangle v_j(\bm{r}',t'), \label{eq:j_corr_real}
\end{multline}
where $k_B$ is the Boltzmann's constant, $T$ the system's temperature, and the correlation function, $\langle \cdot \rangle$, is defined for a pair of general phase-space variables, $\hat X$ and $\hat Y$, as the equilibrium expectation over the initial conditions of a molecular trajectory, $\Gamma_t$, of the time-lagged product of the values of the variables:
\begin{equation}
  \begin{aligned}
    \langle \hat X(t) \hat Y(t') \rangle &= \langle \hat X(t-t') \hat Y(0) \rangle \\ &= \int X(\Gamma_{t-t'}) Y(\Gamma_0) \rho(\Gamma_0) d\Gamma_0.
  \end{aligned}
  \label{eq:corrXY}
\end{equation}
The dependence of the correlation function in Eq.~\eqref{eq:corrXY} on the time difference is due to time-translation invariance ensuing from the equilibrium condition. By the same token, space-translation invariance makes the correlation function in Eq.~\eqref{eq:j_corr_real} only depend on $\bm{r}-\bm{r}'$, turning the integral in $d\bm{r}'$ into a convolution. By using the continuity equation to replace the time derivative of the density with the divergence of the conjugate current density, Eq.~\eqref{eq:j_corr_real} can be cast into a linear relation between the Fourier transforms of the longitudinal component of the current density and the forces acting on the system:
\begin{align}
      \tilde{\jmath}_{\parallel i}(\bm{k},t) 
      =-\sum_j \int_{-\infty}^t \chi_{\parallel ij}(\bm{k},t-t') \tilde{f}_{\parallel j}(\bm{k},t') dt',
    \label{eq:j_corr}
\end{align}
where $\tilde g_\parallel(\bm{k})=\frac{1}{k}\bm{k}\cdot \tilde{\bm{g}}(\bm{k})$ indicates the Fourier transform of the longitudinal component of a generic vector field $\bm{g}(\bm{r})$, $\chi_{\parallel ij}(\bm{k},t)=\frac{1}{k_BT}\bigl\langle \tilde{\hat{\jmath}}_{\parallel i}(\bm{k},t) \tilde{\hat{\jmath}}_{\parallel j}(-\bm{k},0) \bigr\rangle$ is the longitudinal susceptibility of the current densities, and $\bm{f}_i(\bm{r})=-\nabla v_i(\bm{r})$ is the force field associated with the $v_i$ perturbing potential. For the sake of streamlining the notation and without much loss of generality, we will restrict ourselves to longitudinal perturbations and response currents, and drop the ``$\parallel$'' suffix from currents and forces. If the external perturbation is independent of time, in the long-wavelength limit Eq. \eqref{eq:j_corr} results in the Onsager relation between particle fluxes and applied forces \cite{Onsager1931a,*Onsager1931b}:
\begin{align}
    J_n = \sum_{nm} \Lambda_{nm}F_m, \label{eq:Onsager}
\end{align}
where the index $n=(i,\alpha)$ denotes for short the combination of the indices $i$, for the conserved charge, and $\alpha$, for the Cartesian component; the \textit{fluxes} $J_n=\frac{1}{\Omega} \tilde{\jmath}_n(0)$ and \textit{forces} $F_n=\frac{1}{\Omega} \tilde{f}_n(0)$ are the macroscopic averages of the current densities and force fields, respectively; $\Omega$ is the system's volume, and
\begin{align}
  \Lambda_{nm}=\frac{\Omega}{k_BT}\int_0^\infty \Bigl \langle \hat{J}_n(t) \hat{J}_m(0) \Bigr \rangle dt \label{eq:OGK}
\end{align}
is the matrix of Onsager's transport coefficients \cite{Onsager1931a,*Onsager1931b}. In the case of a charged fluid, for instance, the steady state charge flux, $\bm{J}$, induced by a stationary electric field, $\bm{E}$, is given by: $\bm{J}=\sigma \bm{E}$, where the static electrical conductivity is $\sigma=\frac{\Omega}{k_BT}\int_0^\infty \bigl \langle \hat{\bm{J}}(t) \cdot \hat{\bm{J}}(0) \bigr \rangle dt$.

\begin{figure}
    \centering
    \includegraphics[width=0.95\columnwidth]{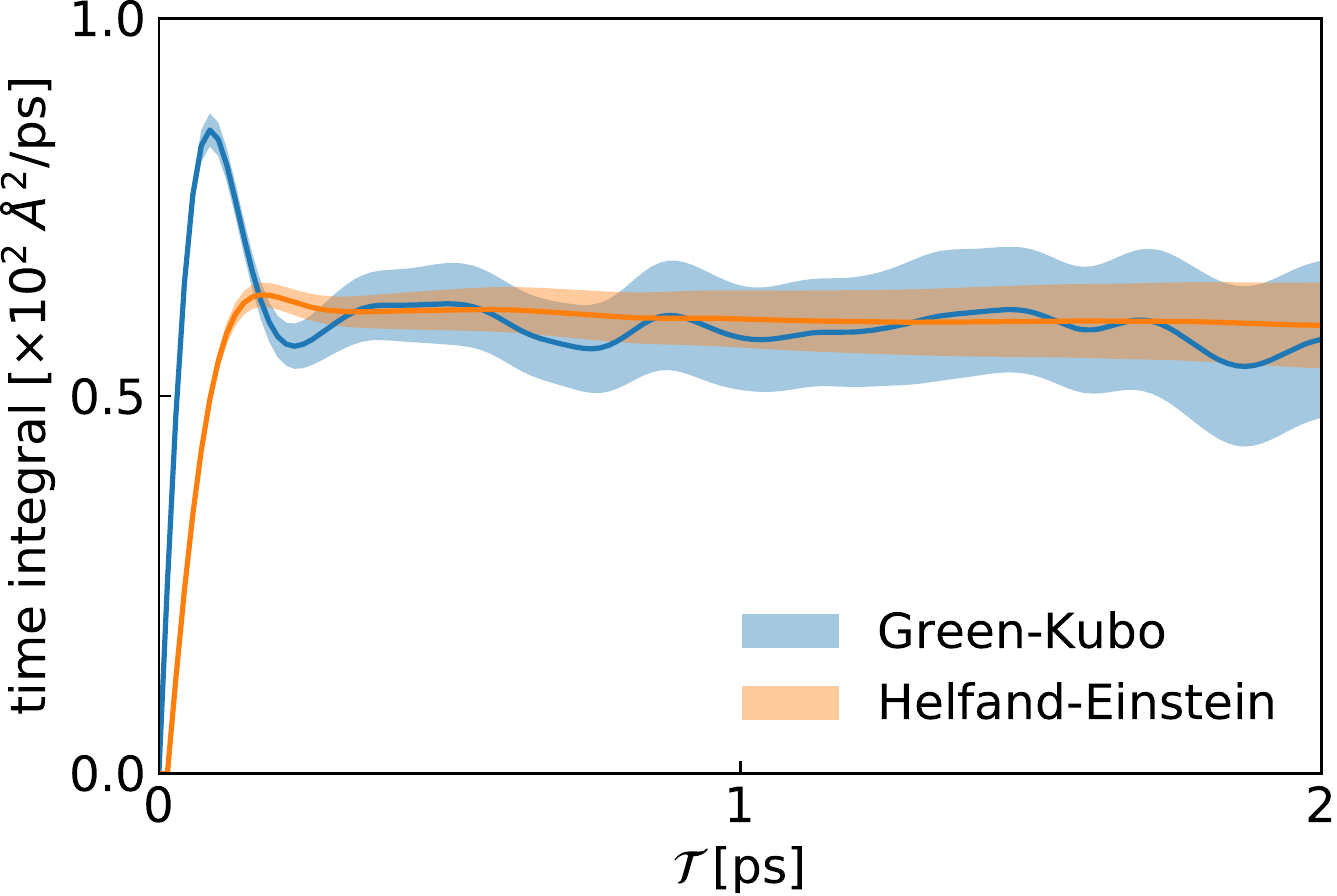}
    \caption{Comparison between the Green-Kubo and the Helfand-Einstein integrals, see Eq. \eqref{eq:Einstein-Helfand}, of the autocorrelation function of the charge flux for an \textit{ab initio} MD simulation of molten KCl. The shaded area represents the confidence interval estimated via standard block analysis. }
    \label{fig:GKHE}
\end{figure}

We can reformulate the GK expression of Onsager's coefficients in another equivalent representation, the so-called Helfand-Einstein (HE) formula, which will be expedient in the following and is also better behaved statistically, based on the identity:
\begin{equation}
    \int_0^\upt dt \int_0^\upt dt' \, f(t' -t)
    = 2\upt \int_0^\upt dt \, \left( 1 - \frac{t}{\upt} \right) f(t){,}
\end{equation}
{valid for any even function, $f(-t)=f(t).$}
%
Let $\hat{\bm{J}}(t)$ be a \emph{stationary} stochastic process representing a conserved flux, so that
$f(t,t') = \langle \hat{\bm{J}}(t)~\cdot~ \hat{\bm{J}}(t')\rangle$ only depends upon $|t-t'|$. By applying the identity above, we obtain:
\begin{equation}
    \int_0^\infty \left \langle \hat{\bm{J}}(t) \cdot \hat{\bm{J}}(0) \right \rangle dt = \lim_{\upt\to\infty} \frac{1}{2\upt} \left \langle \left | \int_0^\upt \hat{\bm{J}}(t) dt \right |^2 \right \rangle. \label{eq:Einstein-Helfand}
\end{equation}
This is called the HE formula, which gives a transport coefficient of some conserved charge, as the ratio of the mean-square \emph{dipole}, $\bm{D}(\upt)=\int_0^\upt\bm{J}(t)dt$, displaced by the conserved flux, $\bm J$, in a time $t$ and time itself. This argument was first exploited by Einstein in his celebrated paper on Brownian motion \cite{Einstein1905} to establish the relation between diffusivity and velocity auto-correlation functions, and later extended by Helfand to general transport phenomena \cite{Helfand1960}.
A comparison between the numerical performance of the GK and the HE formulas is displayed in Fig. \ref{fig:GKHE} in the case of charge transport in a molten salt. The better stability of the HE integral is evident: not only does the HE integral converge faster than the GK one, but the variance on the first, even though growing linearly with time, is much smaller than the one on the second.

\begin{figure}
    \centering
    \includegraphics[width=0.95\columnwidth]{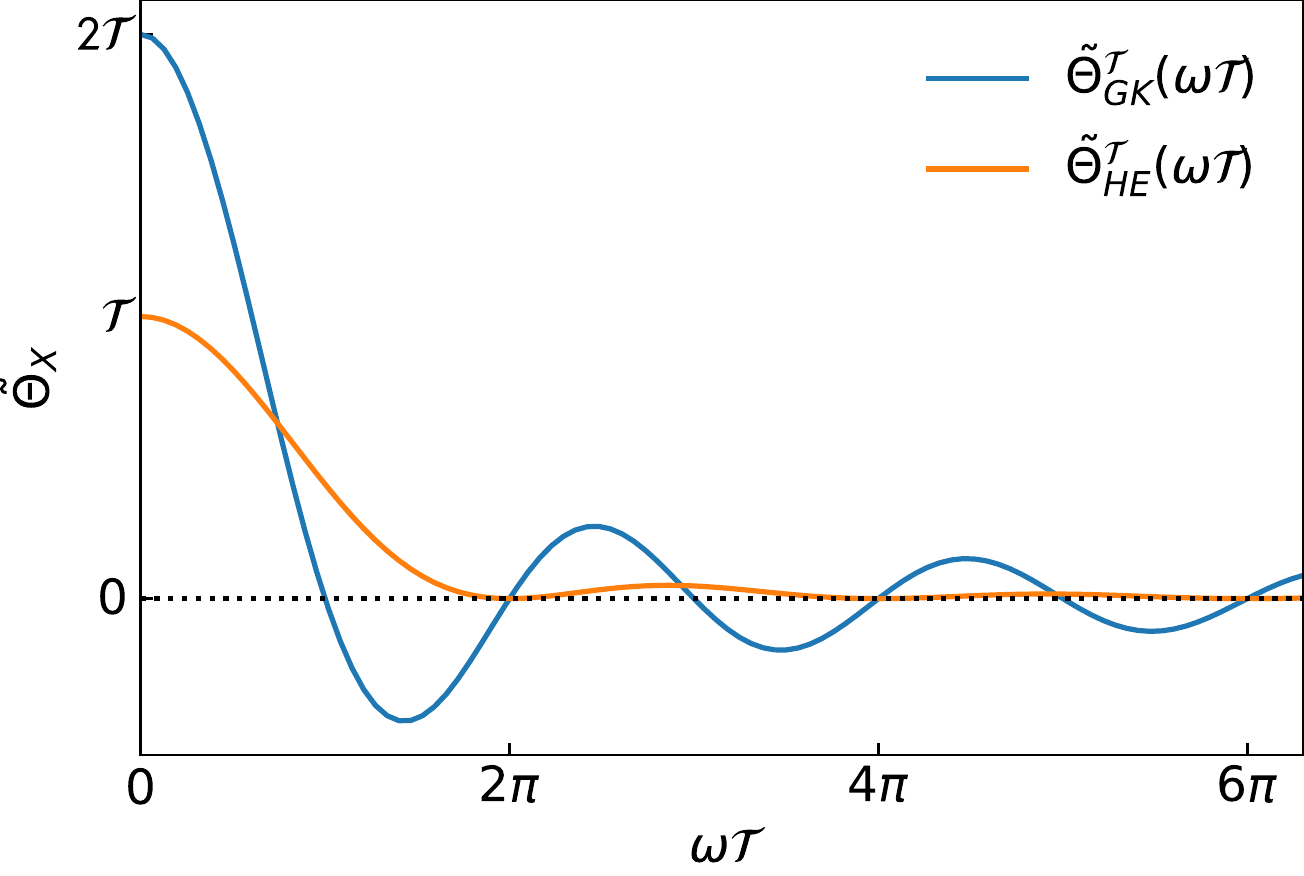}
    \caption{Fourier transforms of the cutoff functions entering the finite-time GK and HE expressions for the transport coefficients as integrals over the entire real axis, see Eq. \eqref{eq:Theta(t)}. }
    \label{fig:ThetaCutoff}
\end{figure}

In order to understand the better statistical behaviour of the HE representation of transport coefficients with respect to the GK one, we leverage the relation between conductivities and the spectral properties of the conserved fluxes, which will also be instrumental in our subsequent considerations on data analysis in Sec. \ref{sec:cepstral}. Let us define:
\begin{equation}
    \begin{aligned}
        \lambda &= \lim_{\upt\to\infty} \lambda_{GK}(\upt) \\
        &= \lim_{\upt\to\infty} \lambda_{HE}(\upt) \\
        &= \frac{1}{2}S(0), 
    \end{aligned}\label{eq:lambdaSmezzi}
\end{equation}
where
\begin{equation}
    \begin{aligned}
    \lambda_{GK}(\upt)&=\int_0^\upt \left \langle \hat{\bm{J}}(t) \cdot \hat{\bm{J}}(0) \right \rangle dt, \\
     \lambda_{HE}(\upt)&= \int_0^\upt \left \langle \hat{\bm{J}}(t) \cdot \hat{\bm{J}}(0) \right \rangle \left ( 1-\frac{t}{\upt} \right ) dt, \\
    S(\omega)&=\int_{-\infty}^\infty \left \langle \hat{\bm{J}}(t) \cdot \hat{\bm{J}}(0) \right \rangle \mathrm{e}^{i\omega t}dt.
    \end{aligned}
\end{equation}
One has:
\begin{equation}
    \begin{aligned}
        \lambda_{GK}(\upt) &= \frac{1}{2}\int_{-\infty}^\infty \Theta_{GK}^\upt(t) \left \langle \hat{\bm{J}}(t) \cdot \hat{\bm{J}}(0) \right \rangle dt   \\
        \lambda_{HE}(\upt) &= \frac{1}{2}\int_{-\infty}^\infty \Theta_{HE}^\upt(t) \left \langle \hat{\bm{J}}(t) \cdot \hat{\bm{J}}(0) \right \rangle dt,
    \end{aligned}
\end{equation}
where
\begin{equation}
    \begin{aligned}
        \Theta_{GK}^\upt(t) &= \begin{cases} 1 & \text{for} ~ |t|\le \upt \\ 0 & \text{otherwise} \end{cases}, \\
        \Theta_{HE}^\upt(t) &= \begin{cases} 1 - {\frac{\displaystyle |t|}{\displaystyle \upt}}& \text{for} ~ |t|\le \upt \\ 0 & \text{otherwise} \end{cases} .
    \end{aligned}
    \label{eq:Theta(t)}
\end{equation}
By using the Parseval-Plancherel identity \cite{Parseval}, one gets:
\begin{equation}
  \lambda_X(\upt) = \frac{1}{4\pi} \int_{-\infty}^{\infty} \tilde \Theta_X^\upt(\omega) S(\omega) d\omega, \label{eq:lambda_X}
\end{equation}
where $X=GK\text{ or }HE$ and $\tilde \Theta_X^\upt(\omega) = \int_{-\infty}^\infty \Theta_X^\upt(t) \mathrm{e}^{i\omega t}dt $. The two functions
\begin{equation}
    \begin{aligned}
        \tilde{\Theta}^\upt_{GK}(\omega) &= 2 \upt \mathrm{sinc}(\omega \upt) \\
        \tilde{\Theta}^\upt_{HE}(\omega) &= \upt \mathrm{sinc}^2(\tfrac{\omega \upt}{2}) ,
    \end{aligned}
\end{equation}
where $\mathrm{sinc}(x) \doteq \sin(x)/x$ is the cardinal sine function, are displayed in Fig.~\ref{fig:ThetaCutoff}: as $\upt\to\infty$, they both tend to $2\pi \delta(\omega)$, where $\delta$ indicates the Dirac delta function. One sees that the statistical accuracy of the HE estimate of the transport coefficients is higher than GK's. Using the findings of Sec. \ref{sec:cepstral}, in Appendix \ref{App:A} we demonstrate that, in the large-$T$ limit, one has indeed the following relation between the statistical uncertainties: $\Delta \lambda_{HE} \approx \frac{1}{\sqrt{3}} \Delta \lambda_{GK}$.

\section{Gauge invariance} \label{sec:gauge}

{\subsection{General theory}}
The gauge-invariance principle of transport coefficients is the condition by which transport coefficients are largely insensitive to the specific definition of the fluxes. In fact, from a microscopic standpoint, any two conserved densities, $\hat{q}'(\bm{r},t)$ and $\hat{q}(\bm{r},t)$, whose integrals over a volume $\Omega$ differ by a quantity that scales as the volume boundary, should be considered equivalent in the large volume limit, $\Omega \to \infty$. For instance, two equivalent densities may differ by the divergence of a (bounded) vector field $\hat{\bm{b}}(\bm{r},t)$:
\begin{equation}
  \hat{q}'(\bm{r},t)=\hat{q}(\bm{r},t) - \nabla\cdot \hat{\bm{b}}(\bm{r},t). \label{eq:gauge_transformation}
\end{equation}
In this sense, $\hat{q}(\bm{r},t)$ and $\hat{q}'(\bm{r},t)$ can be thought of as different \emph{gauges} of the same scalar field. Since $\hat{Q}=\int_\Omega \hat{q}(\bm{r},t) d\bm{r}$ is also conserved, for any given gauge of the conserved density, $\hat{q}(\bm{r},t)$, a conserved current density can be defined, $\hat{\bm{\jmath}}(\bm{r},t)$, so as to satisfy the continuity equation, Eq.~\eqref{eq:continuity}. By combining Eqs.~\eqref{eq:gauge_transformation} and \eqref{eq:continuity} we see that conserved current densities, as well as macroscopic fluxes, transform under a gauge transformation as:
\begin{equation}
\begin{aligned}
 \hat{\bm{\jmath}}'(\bm{r},t) & = \hat{\bm{\jmath}}(\bm{r},t) + \dot{\hat{\bm{b}}}(\bm{r},t), \\ 
  \hat{\bm{J}}'(t) & = \hat{\bm{J}}(t) + \dot{\hat{\bm{B}}}(t), 
\end{aligned}
\end{equation}
where $\hat{\bm{B}}(t)=\frac{1}{\Omega} \int \hat{\bm{b}}(\bm{r},t)d\bm{r}$, the time dependence being \textit{implicit} via the phase-space point $\Gamma_t$, \emph{e.g.} $\hat{\bm J}(t)=\bm{J}(\Gamma_t)$, and where the dot indicates the Poisson brackets with the \emph{unperturbed} Hamiltonian. We conclude that the macroscopic energy fluxes in two different energy gauges differ by the total time derivative of a bounded phase-space vector function. Nonetheless, both definitions \textit{must} lead to the same transport coefficient. To prove such a \textit{gauge-invariance} principle, let us consider a generic transport process with conserved flux represented by the stochastic stationary process $\hat{\bm{J}}(t)$, and define the (generalised, conserved) \textit{charge displacement} per unit volume $ \hat{\bm{D}}(\upt)=\int_0^{\upt} \hat{\bm{J}}(t) \, dt$. According to the HE formulation we can express the transport coefficient of the process, $\sigma$, as
\begin{align}
\sigma &= c \lim_{\upt\to\infty} \frac{\langle | \hat{\bm{D}}(\upt) |^2 \rangle}{2\upt},
\end{align}
where the factor $c$ is specific to the transport process considered. The addition of a time-bounded term $\hat{\bm{B}}(\upt)$ to $ \hat{\bm{D}}(\upt)$, resulting in the new displacement $ \hat{\bm{D}}^\prime(\upt) \equiv  \hat{\bm{D}}(\upt) + \hat{\bm{B}}(\upt)$ produces a transport coefficient $\sigma'$ which coincides with $\sigma$.
In fact, by direct calculation, we have
\begin{align}
    \sigma' &\equiv c \lim_{\upt\to\infty} \frac{\langle |\hat{\bm{D'}}(\upt) |^2 \rangle}{2\upt} \nonumber \\
    &=c \lim_{\upt\to\infty} \frac{
    \overbrace{\langle |\hat{\bm{D}}(\upt) |^2 \rangle}^{\mathcal{O}(\upt)} + \overbrace{2 \langle \hat{\bm{B}}(\upt)  \hat{\bm{D}}(\upt)  \rangle}^{\mathcal{O}(\upt^{1/2})} +
    \overbrace{\langle | \hat{\bm{B}}(\upt) |^2 \rangle}^{\mathcal{O}(\upt^0)}}{2\upt} \nonumber \\&=c \lim_{\upt\to\infty} \frac{\langle |\hat{\bm{D}}(\upt) |^2 \rangle}{2\upt} \equiv \sigma.
\end{align}

\subsection{Some considerations on boundary conditions}\label{sec:considerationsPBC}

To conclude this section we spend a few words to examine the role of boundary conditions (BC) in practical molecular dynamics simulations of a finite sample of the system. First of all, the simulation box must be larger than the relevant correlation/diffusion lengths of the system, in order for equilibrium properties to be independent of specific BC adopted in the simulation. In EMD simulations, periodic BC (PBC) are preferred since \textit{i)} they minimize size effects, and \textit{ii)} the limit $\lim_{\upt\to\infty} \int_0^\upt \langle \hat{\bm{J}}(t)  \hat{\bm{J}}(0) \rangle \, dt$ commutes with thermodynamic limit, where $L,N\to\infty$ while the density $L^3/N$ is kept fixed. This commutation no longer holds in open BC (OBC) at thermodynamic equilibrium, where the asymptotic time limit, $\upt\to\infty$, must be taken only after the thermodynamic limit is performed.

Differently stated, PBC have to be preferred with respect to OBC, since they are the only ones which can sustain a steady-state flux \cite{RestaJuelich}. Nonetheless, this poses some issues in the definitions of the fluxes, since the textbook definition relying on the first moment of the time-derivative of the (periodic) charge density $\hat{q}(\bm{r},t)$,
\begin{equation}
    \hat{\bm{J}}(t) = \frac{1}{\Omega} \int_{\Omega} \dot{\hat{q}}(\bm{r},t) \, \bm{r} d\bm{r}, \label{eq:J_t}
\end{equation}
cannot be employed since $\bm{r}$ is ill-defined in extended, PBC-closed systems. 
{Strictly speaking, Eq.~\eqref{eq:J_t} is ill-defined in PBC for the very same reason why macroscopic polarisation is so in insulators \cite{resta1993macroscopic}. The formal meaning of this equation is that it should be considered as the leading order of a Taylor series of the Fourier transform of the time derivative of the conserved density in powers of its argument: $\dot{\hat{\tilde q}}(\bm k,t) =-i \bm k \cdot \hat{\bm J}(t)+\mathcal{O}(k^2)$. Computer simulations performed for systems of any finite size, $L$, give access to the Fourier components of conserved (current) densities at finite wave-vectors whose minimum magnitude is $|{\bm k}_{min}|=\frac{2\pi}{L}$. Let $c_L(t)= \bigl \langle \hat{\tilde{\bm\jmath}}({\bm k}_{min},t) \cdot \hat{\tilde{\bm\jmath}}(-{\bm k}_{min},0) \bigr \rangle_L $ be the (spatial Fourier transform of the) current-current correlation function evaluated at $ \bm k = {\bm k}_{min}$, which can be easily evaluated from a MD simulation. Unfortunately, $c_L(t)$ cannot be directly used to estimate, not even by extrapolation, the values of the transport coefficients, because the corresponding GK integral vanishes for any finite system size. In practice, the flux to be used to evaluate transport coefficients from the GK formula is obtained from Eq.~\eqref{eq:J_t} by formal manipulations that make the flux boundary insensitive. The time correlation function $C_L(t) \doteq \langle \hat{\bm{J}}(t) \cdot \hat{\bm{J}}(0) \rangle_L$, computed for a system of size $L$, is well-defined as well, and has the property that, for any given $t$,
\begin{equation}
    \lim_{L\to \infty} \bigr ( C_L (t) - c_L(t) \bigr)=0  . \label{eq:C-conv}
\end{equation}
The tricky thing is, the convergence of the limit in Eq.~\eqref{eq:C-conv} is not uniform and, while $\int_0^\infty c_L(t)dt=0$, $\int_0^\infty C_L(t)dt$ is the non-vanishing GK integral yielding the transport coefficient we are after.
}
Therefore, definitions for the fluxes suitable to PBC must be properly designed not only from the speculative standpoint, but also to run meaningful simulations. In what follows, we shall largely use these PBC-based definitions together with the gauge invariance principle to draw general conclusions on transport properties which do not depend on the system size, and hold in the thermodynamic limit.
\section{Convective invariance} \label{sec:convective}
In a multicomponent system the (relevant) conserved charges are the energy and the particles number (or, equivalently, the masses) of \emph{each} atomic species. 
Since the total-mass flux (\emph{i.e.} the total linear momentum) is itself a constant of motion, for a $K$-species system the number of \emph{independent} conserved fluxes is equal to $K$ (energy, plus $K-1$ partial masses) \cite{Foster1975}.
Further constraints may reduce the number of relevant conserved fluxes. For instance, energy flux becomes the only relevant conserved flux in solids or in one-component molecular liquids, as long as the molecules do not dissociate. 
In true multicomponent systems (molten salts, solutions, etc.), neither the mass fluxes of the single atomic species are constant of motion nor their integral is a bound quantity.
This fact must be taken into account in practical simulations of heat transport, since the thermal conductivity relates, by definition, a gradient of temperature to the induced energy flux \textit{in the absence of convection}, i.e. when {the non equilibrium average of} the macroscopic mass flux vanishes.

To make things simpler but also more quantitative, let us consider a two-component system, like, \textit{e.g.}~a molten salt. The conserved fluxes are the energy flux $\bm J_E$ and the mass flux of one of the species $\bm J_M$.
The following system of phenomenological equations hold:
\begin{equation}
\begin{cases}
\bm J_E &= \Lambda^{EE} \nabla\left(\frac{1}{T}\right) + \Lambda_{EM} \nabla\left(\frac{\Delta\mu}{T}\right) \\
\bm J_M &= \Lambda_{ME} \nabla\left(\frac{1}{T}\right) + \Lambda_{MM} \nabla\left(\frac{\Delta\mu}{T}\right)
\end{cases}
\end{equation}
where $\Delta \mu$ is the difference between the chemical potentials of the two species \cite{Sindzingre1990}, and where $\bm J_{M}\equiv \bm J_{M_1} = - \bm J_{M_2}$, $\bm J_{M_l}$ being the mass flux of species $l$, the last step following from the conservation of linear momentum.
When we set the mass flux $\bm J_M = 0$, 
\begin{equation}
\nabla\left(\frac{\Delta\mu}{T}\right) = - \frac{\Lambda_{ME}}{\Lambda_{MM}} \nabla\left(\frac{1}{T}\right)
\end{equation}
which can be substituted in the first equation of the system to finally find
\begin{align}
\bm J_E &= -\kappa \nabla T \\
\kappa &\equiv \frac{1}{T^2} \left[ \Lambda_{EE} - \frac{(\Lambda_{EM})^2}{\Lambda_{MM}} \right] \label{eq:kappa2comp}
\end{align}
where we employed the symmetric property $\Lambda_{EM}=\Lambda_{ME}$. In the light of GK theory, 
the thermal conductivity $\kappa$ is obtained by removing from the GK integral of the energy flux a term which represents the contributions of convection/mass diffusion to heat flow \cite{Galamba2007}. 

It is straightforward to verify that a change in the definition of the microscopic energy flux by any multiple of the mass flux,
\begin{equation}
\hat{\bm J}_{E}^\prime = \hat{\bm J}_E + c \hat{\bm J}_M, \quad c \in \mathbb{R},
\end{equation}
does not affect $\kappa$, even if such change \textit{does change} each of the Onsager coefficients in Eq.~\eqref{eq:kappa2comp}. We dub this peculiar property the ``convective invariance'' principle.
This can be easily extended to more than two species, $K>2$, thanks to standard linear algebra techniques. In such case, $\kappa$ becomes
\begin{equation}
\kappa = \frac{1}{T^2} \left(\Lambda_{EE} - \sum_{l,m=1}^{K-1} \Lambda_{EM_l} (\mathrm{L}^{-1})^{l m} \Lambda_{EM_m}  \right)
\end{equation}
where $\mathrm{L}=\{\Lambda^{M_l M_m}\}$ is the square matrix of Onsager coefficients of the mass fluxes.
The convective-invariance principle reads
\begin{equation}
\hat{\bm J}_{E}^\prime = \hat{\bm J}_E + \sum_{l=1}^{K-1} c_l \hat{\bm J}_{M_l}, \quad c_l \in \mathbb{R} \quad\Rightarrow \quad \kappa'=\kappa
\end{equation}
\textit{Any linear combination of the mass fluxes can be added to the energy flux without affecting the thermal conductivity}. This statement has a direct, crucial consequence concerning \textit{ab initio} calculations: the heat conductivity cannot in fact depend on whether atomic cores contribute to the definition of the atomic energy, as they would in an all-electron calculation, or not, as they would when using pseudo-potentials. In the latter case, the energy of isolated atoms would depend on the specific form of pseudo-potential adopted, which is to a large extent arbitrary, while the heat conductivity in all cases should not. Thanks to convective invariance, shifting the zero of energy of each species by a quantity $\delta E_l$ would result in a change $\hat{\bm J}_{E}^\prime = \hat{\bm J}_E + \sum_{l=1}^{K-1} \frac{\delta E_l}{M_l} \hat{\bm J}_{M_l}$ which does not affect $\kappa$, just like physical intuition would suggest. Finally, from a more practical way, convective invariance also avoids the calculation of partial enthalpies to dispose of the spurious self-energy effects, a rather tedious and cumbersome task \cite{Debenedetti1987,Vogelsang1987,Sindzingre1989}. 

\section{Cepstral analysis} \label{sec:cepstral}
\subsection{Wiener-Khintchin theorem}
Cepstral analysis is a powerful spectral method introduced in the '60s for the analysis of time series, mainly in the field of speech recognition and sound engineering \cite{Bogert1963}. In order to deploy its power to extract the transport coefficient from the time series of the relevant conserved fluxes, we must shift to a Fourier-space representation of stochastic processes, as allowed by the Wiener-Khintchin theorem \cite{Wiener1930,Khintchine1934}. The latter states that the expectation of the squared modulus of the Fourier transform of a stationary process is the Fourier transform of its time correlation function.
We can thus apply this result to the case where the stochastic process is a conserved flux, $\hat{J}(t)$ [the Cartesian indices have been omitted for clarity], and generalize Eq.~\eqref{eq:Einstein-Helfand} to the finite-frequency regime as follows: 
\begin{equation}
  \begin{aligned}
    S_\upt(\omega) &= \frac{1}{\upt} \left \langle \left | \int_0^\upt \hat{J}(t) \mathrm{e}^{i\omega t}dt \right |^2 \right \rangle \\
    &= 2\mathfrak{Re} \int_0^\upt \left \langle \hat{J}(t) \hat{J}(0) \right \rangle \mathrm{e}^{i\omega t}dt + \mathcal{O}(\upt^{-1}).
  \end{aligned}
  \label{eq:Wiener-Khintchine}
\end{equation}
More generally, when several fluxes interact with each other, one can define the \emph{cross-spectrum} of the conserved fluxes as the Fourier transform of the cross time-correlation functions:
\begin{equation}
\begin{aligned}
S_{lm}(\omega) &= \int_{-\infty}^\infty \langle \hat{J}_l(t) \hat{J}_m(0) \rangle \,\mathrm{e}^{i\omega t} dt \\
    &= \frac{1}{\upt} \mathfrak{Re} \left\langle \int_0^\upt \hat{J}_l(t) \mathrm{e}^{-i\omega t}dt \int_0^\upt \hat{J}_m(t) \mathrm{e}^{i\omega t}dt \right\rangle \\ & \quad + \mathcal{O}(\upt^{-1}).
  \end{aligned} \label{eq:Sij(omega)}
\end{equation}
Onsager's coefficients can be thus expressed as:
\begin{equation}
    \Lambda_{lm} =\frac{\Omega}{2 k_B} S_{lm}(\omega=0), \label{eq:GK-S0}
\end{equation}
As we shall see, one can leverage on the Wiener-Khintchin theorem and the gauge invariance principles to obtain good estimates (i.e.~within ~10\% accuracy) of the transport coefficients with relatively short EMD simulations (i.e.~10-100~ps). In practice, this result is based on a particular spectral method named cepstral analysis of time series, which we describe below.

\subsection{Periodograms and power spectra}

\begin{figure}
    \centering
    \includegraphics[width=0.95\columnwidth]{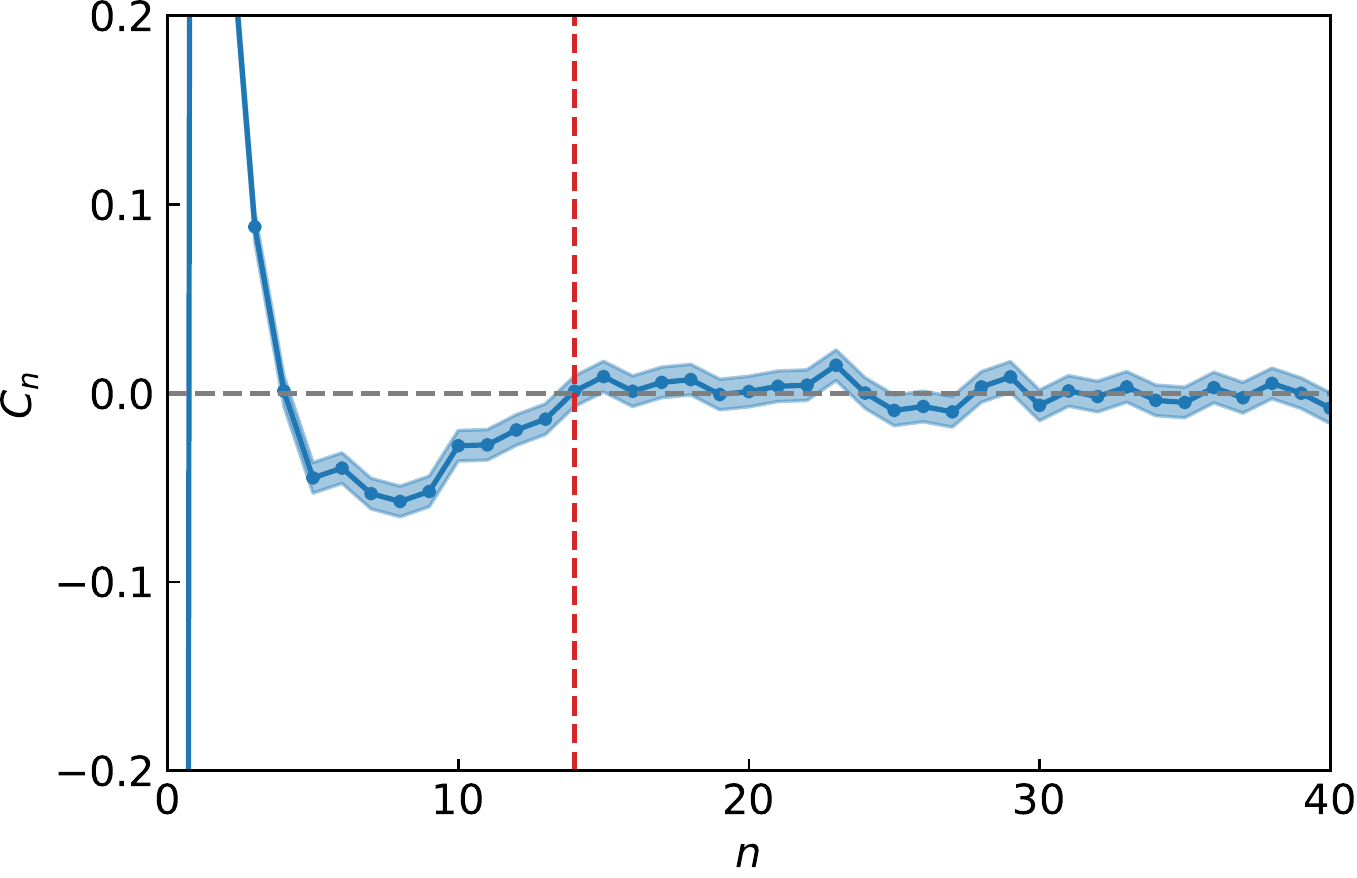}
    \caption{First 40 cepstral coefficients for the log-spectrum of the charge flux in an \textit{ab initio} MD simulation of molten KCl. They are significantly non vanishing only up to $n \approx 15$. The vertical dashed line indicates the number of coefficients to retain according to the AIC.}
    \label{fig:cepstral_coeff}
\end{figure}

Let us focus on one specific flux. In MD simulation we shall have it as a (discrete time) sample, here denoted with the calligraphic font, of the flux process:
\begin{equation}
\stoc{J}_n \equiv \stoc{J}(n\epsilon), \quad n=1,\ldots,N-1
\end{equation}
Here $\epsilon$ is the sampling period, in general a multiple of the timestep of the simulation $\Delta_t$, so that $N\epsilon$ is the total length of the simulation.
The discrete Fourier transform of the flux time series is defined by:
\begin{equation}
  \tilde{\stoc{J}}_{k}=\sum_{n=0}^{N-1} \mathrm{e}^{ 2\pi i\frac{kn}{N}} \stoc{J}_n, \label{eq:Jk}
\end{equation}
for $0 \leq k \leq N-1$.  The \emph{sample power spectrum} $\stoc S_k$, (i.e. the \emph{periodogram}), is defined as
\begin{equation}
\stoc{S}_{k}=\frac{\epsilon}{N} \left |\tilde{\stoc{J}}_{k} \right |^2, \label{eq:periodogram-def}
\end{equation}
and, for large $N$, it is an unbiased estimator of the power spectrum of the process, as defined in Eq.~\eqref{eq:Wiener-Khintchine}, evaluated at
\begin{equation}
  \omega_k=
  \begin{cases}
    2\pi\frac{k}{N\epsilon} & \text{for } k\le\frac{N}{2} \\
    -\omega_{k-\frac{N}{2}} & \text{for } k > \frac{N}{2},
  \end{cases}
\end{equation}
namely:
\begin{equation}
    \langle \stoc S_k \rangle = S(\omega_k).
\end{equation}
Since $\stoc{J}_{n} \in \mathbb{R}$, we have
\begin{equation}
  \tilde{\stoc{J}}_k =\tilde{\stoc{J}}^*_{N-k},
  \stoc S_k=\stoc S_{N-k},
\end{equation}
which, in the continuous limit, amounts to saying that the power spectrum is an even function of frequency: $\stoc{S}(\omega) = \stoc{S}(-\omega)$.
Thanks to this last point, periodograms are usually reported for $0\leq k\leq \frac{N}{2}$ and their Fourier transforms are evaluated as discrete cosine transforms.
The space autocorrelations of conserved currents $\jmath(\bm r,\Gamma(t))$ are usually short-ranged. Therefore, in the thermodynamic limit, the corresponding fluxes $J(\Gamma(t)) = \Omega^{-1} \int_\Omega \jmath(\bm r,\Gamma(t))\, d \bm r$ can be considered sums of (almost) independent identically distributed (iid) stochastic variables: according to the central-limit theorem their equilibrium distribution is Gaussian. Generalizing this argument allows us to conclude that \textit{any conserved-flux process is Gaussian} as well. The flux time series is in fact a multivariate stochastic variable that, in the thermodynamic limit, results from the sum of (almost) independent variables, thus tending to a multivariate normal deviate.
In particular:
\begin{itemize}
\item
  for $k=0$ or $k=\frac{N}{2}$, $\tilde{ \stoc{J}}_k \sim \mathcal{N}\left (0, \frac{N}{\epsilon}S(\omega_k) \right ) \in \mathbb{R}$;
\item
  for \(k\notin\left\{ 0,\frac{N}{2}\right\}\),
  \(\mathfrak{Re}\tilde{\stoc{J}}_k\) and \(\mathfrak{Im}\tilde{\stoc{J}}_k\) are
  independent and both
  \(\sim \mathcal{N}\left (0, \frac{N}{2 \epsilon}S(\omega_k) \right )\)
\end{itemize}
Here $\mathcal{N} (\mu,\sigma^2)$ indicates a normal deviate with mean $\mu$ and variance $\sigma^2$. In conclusion, in the large-$N$ (i.e.~long-time) limit the periodogram of the time series reads:
\begin{equation}
  \stoc{S}_{k} = S \left(\omega_k\right) {\stoc \xi}_{k}, \label{eq:periodogram-distribution}
\end{equation}
the $\stoc{\xi} \sim \frac{1}{2\ell} \chi^2_{2\ell}$ being independent random variables, where $\chi^2_{2\ell}$ is the chi-square distribution with $2\ell$ degrees of freedom and $\ell$ is the number of independent samples of the current (for instance, $\ell=3$ when the 3 equivalent Cartesian components of the flux are considered). In particular, $\bigl \langle\stoc\xi\bigr\rangle=1$ and $\mathrm{var} \bigl( \stoc\xi \bigr) = 1/\ell$. 
Equation~\eqref{eq:periodogram-distribution} shows that $\stoc{S}_{k=0}$ is an unbiased estimator of the zero-frequency value of the power spectrum, i.e.~that $\langle {\stoc S_0} \rangle = S(0)$, and, through Eq.~\eqref{eq:GK-S0}, of the Onsager coefficient we are after.
However, \textit{this estimator is not consistent}, \emph{i.e.} its variance does not vanish in the large-$N$ limit: The multiplicative nature of the statistical noise makes it difficult to disentangle it from the signal.
A way to solve the problem is to apply the \emph{logarithm} to Eq.~\eqref{eq:periodogram-distribution} in order to turn the multiplicative noise into an additive one by defining the \textit{log-periodogram}, $\stoc{L}_{k}$, as:
\begin{equation}
  \begin{aligned}
    \stoc{L}_{k} &= \log \stoc{S}_{k} = \log\left(S(\omega_k) \right) + \log( {\stoc\xi}_k) \\
  \end{aligned} \label{eq:log-PSD}
\end{equation}
The quantities $\log( {\stoc\xi}_k)$ are iid stochastic variables whose statistics is well known: their mean and variance are simply expressed in terms of the digamma and trigamma functions, $\psi(\ell)$ and $\psi^\prime(\ell)$, respectively
\cite{PolyGamma}.
Furthermore, whenever the number of significant (inverse) Fourier components of $\log(S(\omega))$ is much smaller than the length of the time series, applying a low-pass filter to Eq.~\eqref{eq:log-PSD} would result in a reduction of the power of the noise, without affecting the signal. In order to exploit this idea, we define the \emph{cepstrum} of the time series as the inverse Fourier transform of its sample log-spectrum \cite{Childers1977}:
\begin{equation}
   \stoc C_{n} = \frac{1}{N}\sum_{k=0}^{N-1} \stoc L_{k} \mathrm{e}^{-2\pi i\frac{kn}{N}}. \label{eq:sample-cepstrum}
\end{equation}
A generalized central-limit theorem for Fourier transforms of stationary time series ensures that, in the large-$N$ limit, these coefficients
are a set of independent (almost) identically distributed zero-mean normal deviates \cite{Anderson1994,Peligrad2010}.
It also follows that:
\begin{equation}
  \begin{aligned}
     C_n \equiv \langle \stoc{C}_{n} \rangle &= \frac{1}{N}\sum_{k=0}^{N-1} \log\bigl (S(\omega_k) \bigr ) \mathrm{e}^{-2\pi i\frac{kn}{N}},
   \end{aligned} \label{eq:cepstrogram}
\end{equation}
Figure \ref{fig:cepstral_coeff} confirms the typical behaviour of cepstral coefficients calculated from the low-frequency region of the periodogram. We see that only a few coefficients are in fact substantially different from zero, within statistical uncertainty.

\begin{figure}
    \centering
    \includegraphics[width=0.95\columnwidth]{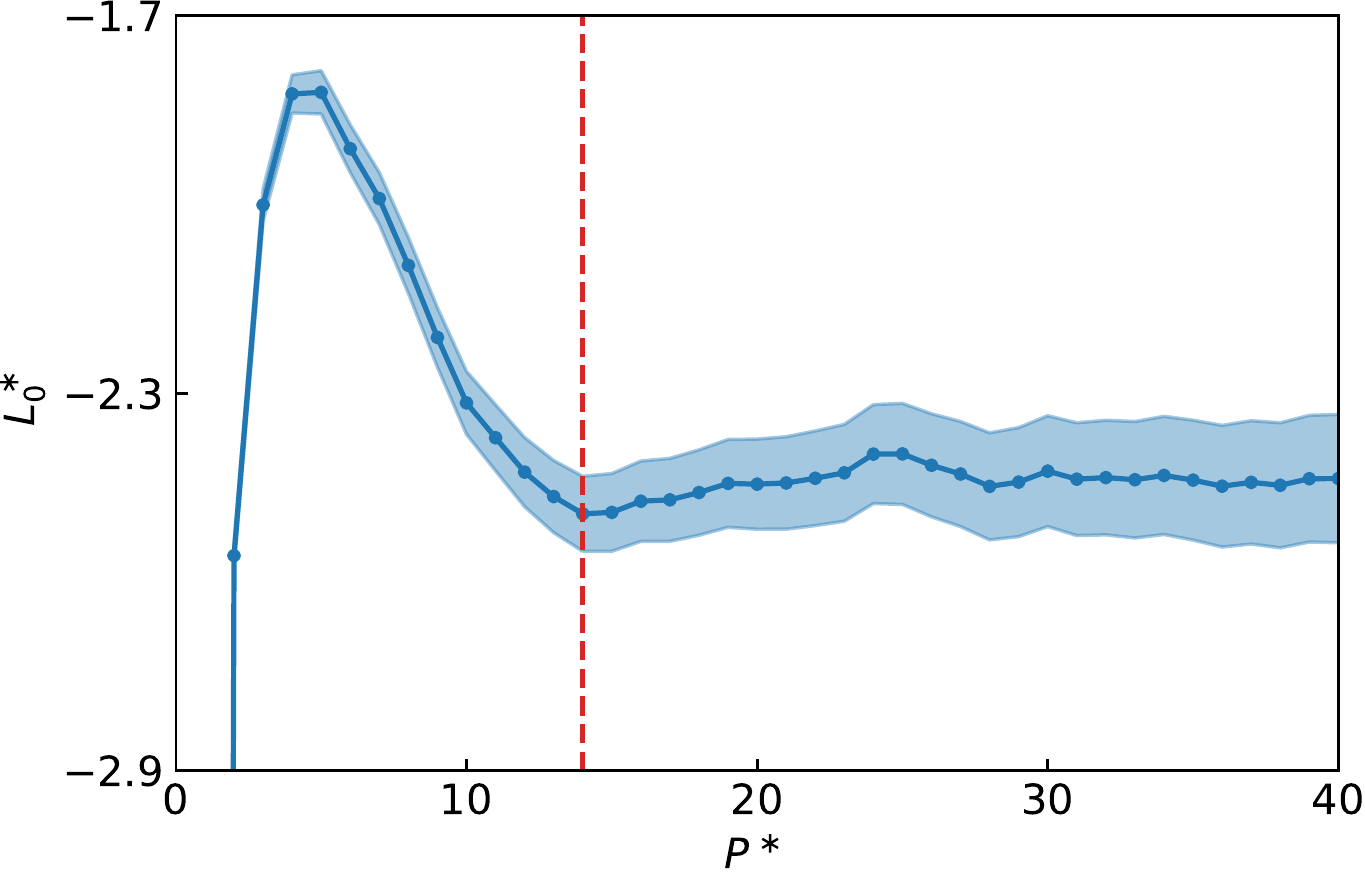}
    \caption{The $\omega=0$ component of the log-spectrum of the charge flux vs $P^\ast$ for an \textit{ab initio} MD simulation of molten KCl. The expectation value (dots, Eq.\eqref{eq:L0*}) and its uncertainty (shaded area, square root of Eq.\eqref{eq:sigma*}) are displayed. The vertical dashed line indicates the optimal $P^\ast$ predicted with the AIC.}
    \label{fig:L0*}
\end{figure}

Let us indicate by $P^*$ the smallest integer such that $C_n \approx 0$ for $P^* \le n \le N-P^*$. By limiting the Fourier transform of the sample cepstrum, Eq.~\eqref{eq:sample-cepstrum}, to $P^*$ coefficients, we obtain an efficient estimator of the zero-frequency component of the log-spectrum, $\stoc{L}_0^*$, whose expectation and variance are
\begin{align}
	L_0^\ast \equiv \langle \stoc{L}_{0}^{*} \rangle &= \log(S_{0}) + \psi(\ell) - \log(\ell), \label{eq:L0*} \\
 	 \text{var}(\stoc{L}_{0}^{*})&=\psi^\prime(\ell)\frac{4P^{*}-2}{N}. \label{eq:sigma*}
\end{align}
We thus see that the logarithm of the Onsager coefficient we are after can be estimated from the cepstral coefficients of the flux time series through Eqs.~(\ref{eq:L0*}-\ref{eq:sigma*}), and that the resulting estimator is always a  \emph{normal} deviate whose variance depends on the specific system \emph{only} through the number of these coefficients, $P^\ast$. Notice that the absolute error on the logarithm of the conductivity directly and nicely yields the relative error on the conductivity itself. The efficacy of this approach obviously depends on our ability to \textit{estimate the number of coefficients necessary} to keep the bias introduced by the truncation to a value smaller than the statistical error, while maintaining the magnitude of the latter at a prescribed acceptable level. In Ref.~\cite{Ercole2017}, it has been proposed to estimate $P^\ast$ using the Akaike's information criterion \cite{Akaike1973}, even if other more advanced \emph{model selection} approaches may be more effective \cite{Claeskens2008}.
A plot of $L_0^\ast$ vs $P^\ast$ is shown in Fig. \ref{fig:L0*}. We immediately realize that $P^\ast$ returned by the AIC  (vertical dashed line) is indeed capable to find the correct value for the log-spectrum---within statistical error---of lowest variance.
Furthermore, thanks to the convective invariance principle described in Sec.~\ref{sec:convective}, the cepstral analysis can be extended to multicomponent systems \cite{Bertossa2019}. This \textit{multivariate} cepstral method has been recently applied to calculate the \textit{ab initio} thermal conductivity of water at planetary conditions from trajectories as short as a few tens of picoseconds \cite{Grasselli2020}. It has been also shown that the multivariate cepstral analysis is able to substantially reduce the statistical error affecting the estimate of thermal conductivity even for one-component systems: it can in fact decorrelate the finite-frequency power spectrum of non-diffusive fluxes (like the mass flux, or the adiabatic electronic flux in ab initio simulations) from the heat flux power spectrum, which will have its total power considerably reduced, and whose low frequency portion will be easier to analyse \cite{Bertossa2019}. The statistical tools for time-series analysis presented in this Section have been implemented in the open-source code \textsc{SporTran}, which is freely downloadable from the GitHub repository \url{https://github.com/lorisercole/sportran} \cite{sportran}.

\section{Ab initio heat transport in insulators} \label{sec:heat}

Until only a few years ago, it was thought that ab initio MD simulations were in general unsuitable to a GK theory of thermal transport, since the continuous electronic density makes any decomposition of the energy flux into local, atomic contributions fully arbitrary \cite{Stackhouse2010b}. This consideration clashes with a reductionist picture whereby a fundamental description of the microscopic interactions should be in principle suitable to describe heat transport in the very general GK theory, as well. Once again, this apparent inconsistency stands on the misconception that the definition of microscopic fluxes must be unique, and it is thus solved thanks to the gauge-invariance principle.
Apart from correcting such a misconception, the gauge-invariance principle proves to be also a rigorous mathematical tool to derive a well-defined expression (out of the infinitely many possibilities!) for the energy flux directly from DFT, with no \textit{ad hoc} approximation tailored on the considered physical system \cite{Carbogno2017}. 
The starting point is the standard DFT definition of the total energy in terms of the Kohn-Sham (KS) eigenvalues $\varepsilon_v$, eigenfunctions $\phi_v(\bm{r})$, and density $n(\bm{r}) = \sum_v |\phi_v(\bm{r})|^2$ \citep{Martin2008}:
\begin{multline}
  E_{\smallDFT} = \frac{1}{2}\sum_{n}M_{n}V_{n}^{2} + \frac{\mathtt{e}^2}{2}\sum_{n,m\ne n}\frac{ Z_{n}Z_{m}}{|\bm{R}_{n}-\bm{R}_{m}|} \\
  + \sum_{v}\varepsilon_{v}-\frac{\mathtt{e}^2}{2}\int\frac{n(\bm{r})n(\bm{r}')}{|\bm{r}-\bm{r}'|}d\bm{r}d\bm{r}'\\
  +\int\left(\epsilon_{\smallXC}[n](\bm{r})-\mu_{\smallXC}[n](\bm{r})\right)n(\bm{r})d\bm{r},
\end{multline}
where $\mathtt{e}$ is the electron charge, $\epsilon_\smallXC[n](\bm{r})$ is a local exchange-correlation (XC) energy per particle defined by the relation $ \int \epsilon_\smallXC[n](\bm{r})n(\bm{r}) d\bm{r}=E_\smallXC [n]$, the latter being the total XC energy of the system, and $ \mu_\smallXC (\bm{r}) = \frac{\delta E_\smallXC }{\delta n(\bm{r})}$ is the XC potential [in this Section and in Sec.~\ref{sec:charge} we drop the hat on top of classical observables to avoid confusion with quantum operators]. From this we can write a DFT energy \textit{density} as \citep{Chetty1992}:
\begin{equation}
  \begin{split}
    E_{\smallDFT} & =  \int e_{\smallDFT}(\bm{r})d\bm{r},\\
    e_{\smallDFT}(\bm{r}) & = e_{el}(\bm{r})+e_{\smallZ}(\bm{r}),
  \end{split}
  \label{eq:DFT-Edensity}
\end{equation}
where:
 \begin{align}
   e_{el}(\bm{r}) & =\mathfrak{Re} \sum_{v}\phi_{v}^{*}(\bm{r})\bigl(H_{\smallKS}\phi_{n}(\bm{r})\bigr) \nonumber \\
   & - \frac{1}{2}n(\bm{r})v_{\smallH}(\bm{r}) +\left(\epsilon_\smallXC (\bm{r}) - \mu_\smallXC  (\bm{r}) \right) n(\bm{r}), \\
   e_{\smallZ}(\bm{r}) & = \sum_{n}\delta(\bm{r}-\bm{R}_{n}) \left(\frac{1}{2}M_{n}V_{n}^{2}+w_{n}\right), \\
   w_{n} & =\frac{\mathtt{e}^2}{2}\sum_{m\ne n}\frac{ Z_{n}Z_{m}}{|\bm{R}_{n}-\bm{R}_{m}|}, \label{eq:DFT-Edensity-breakup}
 \end{align}
$H_\smallKS$ is the instantaneous self-consistent Kohn-Sham Hamiltonian, and $v_\smallH = \mathtt{e}^2 \int d\bm{r}' \frac{n(\bm{r}')}{|\bm{r}-\bm{r}'|}$ is the Hartree potential. An explicit expression for the DFT energy flux,
\begin{equation}
    \bm J^\smallE_\smallDFT = \frac{1}{\Omega} \int \bm{r} \dot{e}_{\smallDFT}(\bm{r}) d\bm{r},\label{eq:illJ}
\end{equation}
obtained by computing the first moment of the time derivative of the energy density in Eqs.~(\ref{eq:DFT-Edensity}-\ref{eq:DFT-Edensity-breakup}), results in a number of terms, some of which are either infinite or ill-defined in PBC, since the position operator is not periodic. Leveraging on gauge invariance and on a careful breakup and refactoring of the various harmful terms, as explained in Refs.~\cite{Marcolongo2014,Marcolongo2016}, we can recast Eq.~\eqref{eq:illJ} in a form suitable to PBC, whose final expression is:
\begin{equation}
  \bm{J}^\smallE_{\smallDFT} =\bm{J}^{\smallH} + \bm{J}^{\smallZ} + \bm{J}^{0} + \bm{J}^{\smallKS} +  \bm{J}^{\smallXC},
\end{equation}
where
\begin{align}
  \allowdisplaybreaks
  &\bm{J}^{\smallH} =
  \frac{1}{4\pi \rOmega \mathtt{e}^2}\int \nabla v_{\smallH}(\bm r) \dot v_{\smallH}(\bm  r) d\bm{r}, \\
  &\bm{J}^{\smallZ} \label{eq:DFT-ionic} =
  \frac{1}{\rOmega} \sum_{n}  \left[\bm{V}_{n}\left(\frac{1}{2}M_{n}V_{n}^{2} + w_{n}\right) \right. \nonumber  \\ 
  & \qquad \qquad\left.  + \sum_{m\ne n}(\bm{R}_{n} - \bm{R}_{m}) \left(\bm{V}_{m} \cdot \frac {\partial w_{n}}{\partial \bm{R}_{m}} \right) \right], \\
  &\bm{J}^{0}  =
  \frac{1}{\rOmega}  \sum_{n} \sum_{v}\left\langle \phi_{v} \left|(\bm{r}-\bm{R}_{n})\left(\bm{V}_{n}\cdot\frac{\partial\hat{v}_{0}}{\partial\bm{R}_{n}}\right)\right| \phi_{v}\right\rangle, \\
  &\bm{J}^{\smallKS} = \label{eq:DFT-KS}
  \frac{1}{\rOmega}  \mathfrak{Re} \sum_{v} \langle \bm{\bar \phi}_v^c | H_{\smallKS}+\varepsilon_v | \dot \phi_v^c \rangle, \\
  &J_{\alpha=x,y,z}^{\smallXC} =
  \begin{cases}
    0 & \mathrm{(LDA)} \\
      -\frac{1}{\rOmega} \int n(\bm{r}) \dot{n}(\bm{r}) \frac{\partial\epsilon^{\smallGGA} (\bm{r})}{\partial(\partial_\alpha n)} d\bm{r} & \mathrm{(GGA)}.
  \end{cases}
\end{align}
Here $\hat v_0$ is the bare, possibly non-local, (pseudo-) potential acting on the electrons and
\begin{equation}
 \begin{aligned}
  |\bm{\bar \phi}_v^c\rangle &= \hat P_c \,\bm{r}  \,|\phi_v \rangle, \\
  |\dot \phi_v^c\rangle &= \dot{\hat P}_v \,|\phi_v\rangle,
\end{aligned} \label{eq:phi-dot}
\end{equation}
are the projections over the empty-state manifold of the action of the position operator over the $v$-th occupied orbital, and of its adiabatic time derivative \citep{Giannozzi2009,Giannozzi2017,giannozzi2020quantum}, respectively, $\hat P_v$ and $\hat P_c = 1 - \hat P_v$ being the projector operators over the occupied- and empty-states manifolds, respectively. Both these functions are well defined in PBC and can be computed, explicitly or implicitly, using standard density-functional perturbation theory \citep{Baroni2001}.
These rather complicated formulae have been implemented in the open-source code \textsc{QEHeat} of the \textsc{Quantum ESPRESSO} project which has been publicly released \cite{Marcolongo2021}.

\section{Ab initio charge transport in ionic conductors} \label{sec:charge}

\subsection{Ab initio charge transport in ionic conductors}
While in metals the current is carried by delocalised conduction electrons, in electronic insulators the electrons are bound to follow adiabatically the ionic motion and no charge transport can occur if the ion positions are frozen. Nonetheless, when ions are allowed to move, like in ionic liquids, charge can be displaced.
Daily life examples range from simple salted water, to liquid electrolytes employed in Li-ion batteries, or to the molten salts (NaCl, KCl, etc.) used as heat exchangers in power plants.\footnote{Pure water itself displays conductive behaviour in its exotic phases at high temperatures/pressures, where H$_2$O molecules are (fully or partially) dissociated. This has implications in both charge \cite{Rozsa2018} and heat transport \cite{Grasselli2020}}
Due to their large electronic bandgap, ionic liquids are in general transparent to visible light and possess a negligible fraction of ``free'', conduction electrons.
When the quantum nature of the electrons is considered, a question arises about a proper definition for the charge flux, $\bm{J}(t)$, to employ in the GK expression for the electrical conductivity, $\sigma$:
\begin{align}
\sigma &= \frac{\Omega}{3k_B T} \int_0^\infty \langle \boldsymbol{{J}}(t) \cdot \boldsymbol{{J}}(0) \rangle \, dt.\label{eq:GKsigma}
\end{align}
Any static partition of the instantaneous total charge in order to express $\bm{J}$ into atomic contributions is in fact totally arbitrary.

The solution comes from the modern theory of polarisation (MTP), which provides a definition for the polarisation $\bm{P}$ valid for extended systems \cite{Resta2008, Resta-Vanderbilt2007, Vanderbilt2018}. 
Just like the electronic Hamiltonian and ground state, $\bm P$ depends on time through the nuclear coordinates only: we can thus apply the chain rule and write $\boldsymbol{J}(t)$ as a sum of atomic
contributions
\begin{equation}
    \begin{aligned}
        \boldsymbol{{J}}(t) & = \dot{\bm{P}}(t)
        = \frac{\mathtt{e}}{\Omega} \sum_{i=1}^{N_{at}} \mathrm{Z}^*_{i}(t) \cdot \boldsymbol{V}_i(t)
    \end{aligned}
    \label{eq:jborn}
\end{equation}
where $\bm{V}_i(t)$ is the atomic velocity of the $i$-th atom and $\mathrm{Z}^*_i(t)$ is a time-dependent tensor whose entries, $Z^*_{i\alpha\beta} \equiv \tfrac{1}{\mathtt{e}}\tfrac{\partial P_\alpha}{\partial R_{i\beta}}$, are the derivative of the cell polarisation along the Cartesian direction $\alpha$ with respect to the atomic displacement of atom $i$ along $\beta$.
The \emph{dynamical charges} $\mathrm{Z}_i$ are called ``Born effective-charge tensors'', and can be computed in perturbation theory \cite{Baroni2001}, or by finite force differences when introducing a finite electric field $\bm{E}$ in accordance to the MTP \cite{Umari2002}. 
Born charge tensors are in general strongly dependent on the atomic positions, and they display large fluctuations along a AIMD trajectory, even for strongly ionic systems \cite{Grasselli2019}.  
For these reasons, the charge flux defined in Eq.~\eqref{eq:jborn} is in general fluctuating and does not vanish identically. Therefore, there is no apparent reason for the GK integral in Eq.~\eqref{eq:GKsigma} to be exactly zero. How come, then, that the electrical conductivity of, say, pure water is zero? And why, instead, under the same microscopic formalism, does a salted water solution have a non-vanishing $\sigma$?
Another strange ``coincidence'' reported in the literature \cite{French2011} is that by replacing in Eq.~\eqref{eq:jborn} the time-dependent Born charge tensors with the predefined constant integer charges---the oxidation numbers of the atoms---to define a new flux 
\begin{equation}
    \bm{J}'(t) = \frac{\mathtt{e}}{\Omega} \sum_{i=1}^{N_{at}} q_{S(i)} \bm{V}_i(t) \label{eq:jprime}
\end{equation}
leads to the same GK integral, Eq.~\eqref{eq:GKsigma}, that is obtained via the alternative definition $\bm J$ in Eq.~\eqref{eq:jborn} \footnote{It must be also remarked the the use of other definitions for the atomic charges, like Bader charges, lead to a wrong electrical conductivity \cite{French2011}.}.

It seems therefore that two main aspects must be understood to answer the fundamental questions raised above: first, the presence of integer numbers suggests some underlying quantisation in the charge transport process; second, the insensitivity of $\sigma$ to the choice of one of the two different definition $\bm{J} \neq \bm{J}'$ hints at a manifestation of the gauge invariance principle of transport coefficients, described in Sec.~\ref{sec:gauge}. We analyse in detail these two aspects in what follows.

\subsection{Quantisation of charge transport}
In 1983, David Thouless showed that for a quantum system in PBC, with non-degenerate ground state evolving adiabatically under a cyclic, slowly varying Hamiltonian $\hat{H}(0) = \hat{H}(\cycle)$, the total displaced dipole, $\Delta \bm \mu$ is quantised, i.e.~the time integral of the current over one cycle $\cycle$ is equal to a triplet of integers, multiplied by the side $L$ of the cell, which we shall assume cubic for simplicity \cite{Thouless1983}:
\begin{equation}
    \Delta \bm \mu = \Omega \int_0^\cycle \bm{J}(t) dt = \mathtt{e} L \bm Q
    \label{eq:Thouless}
\end{equation}
with $\bm Q = (Q_x,Q_y,Q_z) \in \mathbb{Z}^3$.
In AIMD simulations, atomic trajectories are identified by paths in the nuclear configuration space (NCS). The electronic Hamiltonian, $H(\mathbf{R}(t))$, depends parametrically upon time through the nuclear positions, at time $t$, $\mathbf{R}(t) = \left\lbrace \boldsymbol{R}_i(t) \right\rbrace$, with $i = 1, \dots, N_{at}$. A path $\mathcal{C}$ in NCS whose end points are one the the \textit{periodic image} of the other can be thus considered as a cycle for the electronic Hamiltonian. Therefore, we can thus employ Thouless' theorem and show, under very general hypothesis, that the electric dipole displaced along $\mathcal{C}$ is itself partitioned into atomic contributions:
\begin{equation}
    \Delta \bm{\mu}_\mathcal{C} = \mathtt{e} L \sum_{i=1}^{N_{at}} q_S(i) \bm{n}_i, \label{eq:additivity}
\end{equation}
where $\bm{n}_i = (n_{x},n_{y},n_{z})_i$ is the triplet containing the number of cells spanned, along $\mathcal{C}$, by atom $i$ in $x,y$ and $z$ directions, and $q_S(i)$ are integer constants which only depend on the species $S(i)$ of atom $i$, and are shown to coincide with the oxidation numbers suggested by chemical intuition \cite{Grasselli2019}. This provides a quantum-mechanical definition of oxidation numbers where their integerness arises naturally (and not just as an approximation of some real charge), and which identifies them as \textit{intrinsically dynamical} quantities \cite{Jiang2012,Grasselli2019}.  

\subsection{Gauge invariance of electrical conductivity}

\begin{figure}
  \centering
  \includegraphics[scale=0.45]{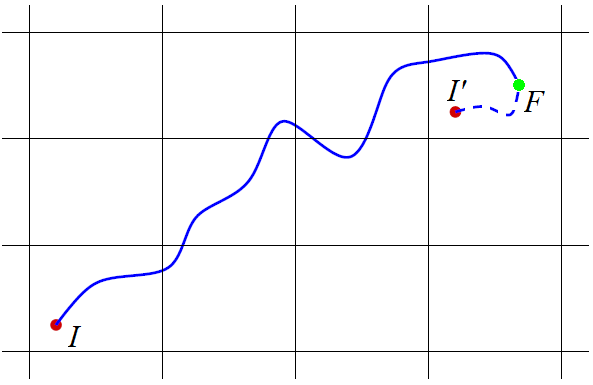}
  \caption{Paths in a two-dimensional NCS representing atomic trajectories. $I'$ is a periodic image of the starting point $I$ belonging to the same periodic cell of the physical endpoint $F$. The path $II'$ is the concatenation of the physical trajectory $IF$ with the path $I'F$ entirely belonging to one cell. }\label{fig:paths}
\end{figure}

\begin{figure}
    \centering
    \includegraphics[width=0.95\columnwidth]{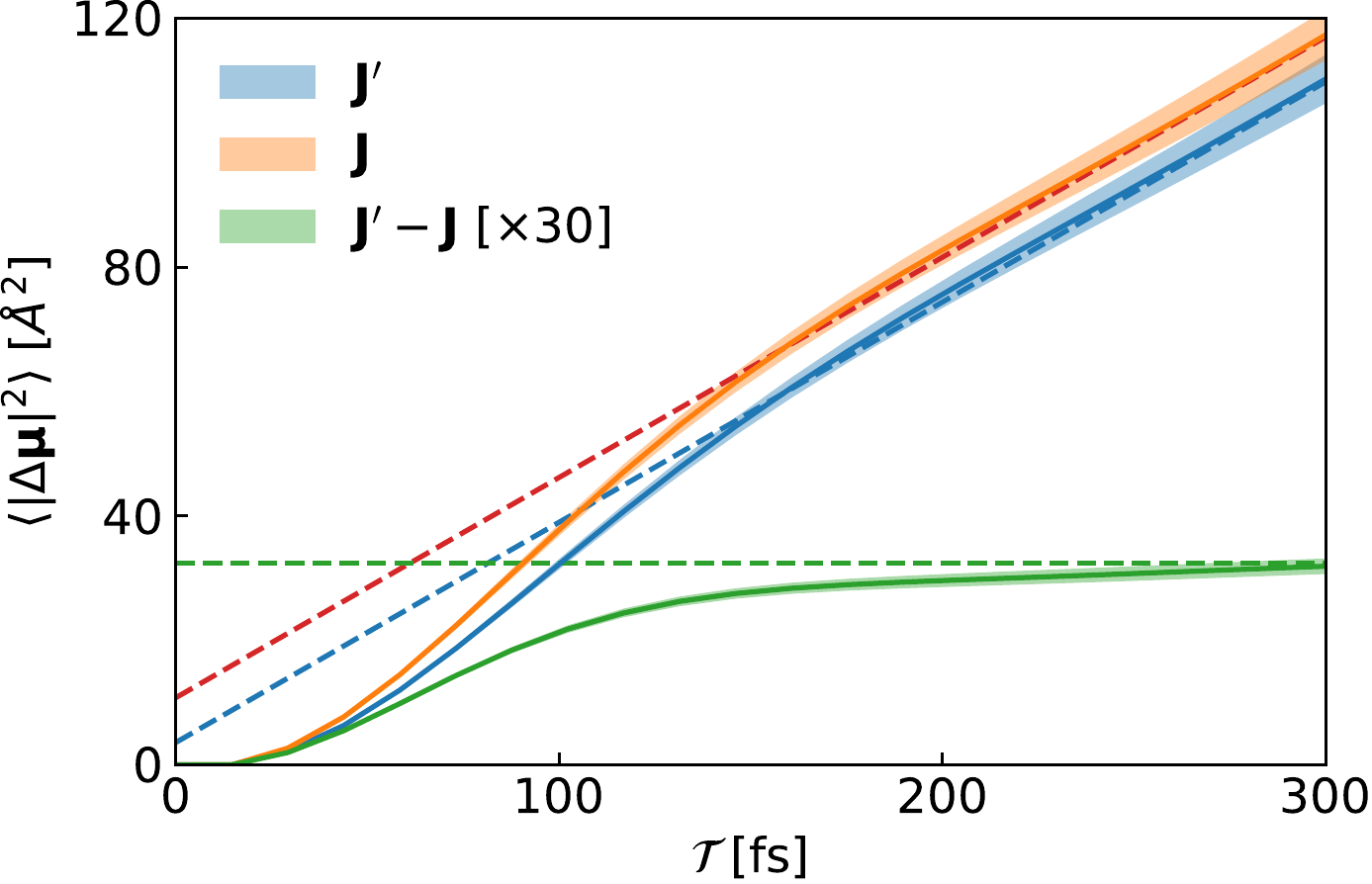}
    \caption{Mean square dipole displaced along a trajectory of molten KCl. The slope of the asymptotic behaviour is proportional to the electrical conductivity. Calculations employing the two different definitions Eq.~\eqref{eq:jborn} and Eq.~\eqref{eq:jprime} share the same slope, and thus the same electrical conductivity. The difference $\bm J' - \bm J$ is instead non diffusive (zero slope). The fits of the curves at asymptotic times are also displayed (dashed lines).}
    \label{fig:slopes}
\end{figure}

We are now ready to combine the theory of quantisation of charge transport and the gauge-invariance principle of transport coefficients to answer the questions raised at beginning of this Section.
Let us consider a physical path in the NCS from an initial configuration $I$ to the configuration $F$, as a result, \textit{e.g.}~of an AIMD simulation. Then, we elongate the path fictitiously up to the point $I'$ which is the replica (periodic image) of the initial point $I$ sharing with the point $F$ the same cell of the nuclear configuration space, as depicted by the dashed line in Fig.~\ref{fig:paths}.
Due to the additivity of integrals, the electric dipole displaced along the physical path $IF$ reads
\begin{align}
\Delta\bm{\mu}_{IF} &\equiv \int_{IF} d\bm{\mu} =   \Delta\bm{\mu}_{II'}  + \Delta\bm{\mu}_{I'F} \;.
\end{align}
Since the open path $I'F$ entirely belongs to one cell, $\Delta\bm{\mu}_{I'F}$ is a bounded quantity.
Therefore, thanks to gauge-invariance, to evaluate $\sigma$ we only need to consider
\begin{equation}
\Delta\bm{\mu}_{II'} = \int_{II'} d\bm{\mu} = \mathtt{e} L \sum_{i=1}^N q_{S(i)} \bm{n}_{i},
\end{equation}
since
\begin{equation}
\sigma \propto \lim_{\upt \rightarrow \infty} \frac{\langle |\Delta\bm{\mu}_{IF}(\upt)|^2 \rangle}{2\upt}
= \lim_{\tau \rightarrow \infty} \frac{\langle \left|\Delta\bm{\mu}_{II'}(\upt)\right|^2 \rangle}{2\upt}. \label{eq:equiv1}
\end{equation}
By the same token, the electric dipole displaced from $I$ to $F$ by to the flux $\bm{J}^\prime(t)$ defined in Eq.~\eqref{eq:jprime} is:
\begin{equation}
\Delta \boldsymbol{\mu^\prime}_{IF}
= \Delta \boldsymbol{\mu}_{II'} +\mathtt{e} \sum_{i=1}^N q_{S(i)} \underbrace{\int_{I'}^F d\boldsymbol{R}_i}_{\text{bounded}} .
\end{equation}
Therefore we can conclude that
\begin{equation}
\begin{split}
  \sigma' \propto \lim_{t \rightarrow \infty} \frac{\langle \left|\Delta\bm{\mu^\prime}_{IF}(\upt)\right|^2 \rangle}{2\upt}
&= \lim_{\upt \rightarrow \infty} \frac{\langle \left|\Delta\bm{\mu}_{II'}(\upt)\right|^2 \rangle}{2\upt} \label{eq:equivalence_sigma}
\end{split}
\end{equation}
which, by comparison with Eq. \eqref{eq:equiv1}, proves the equivalence of the electrical conductivities obtained via Eq.~\eqref{eq:jborn} and Eq.~\eqref{eq:jprime}. This is shown in Fig.~\ref{fig:slopes} for an ab initio MD simulation of molten KCl.

As explained in detail in Ref.~\cite{Grasselli2019}, this result is grounded on the hypothesis, which we dubbed \textit{strong adiabaticity}, that any closed paths in the NCS can be shrunk to a point without closing the electronic gap: this implies---see Eq.~\eqref{eq:additivity}---that charge transport can occur only through a net displacement of the ions, as it is typical of stoichiometric ionic conductors. The breach of strong adiabaticity may instead dictate a \textit{non-trivial} charge-transport regime where charge may be adiabatically transported even in the absence of a net ionic displacement. As shown in Ref.~\cite{Pegolo2020} this non-trivial behaviour is intertwined with the presence, in non-stoichiometric electrolytes, of dissolved yet localised electrons, whose displacement is to a large extent uncorrelated to that of the ions.
{By the same token discussed in Sec.~\ref{sec:considerationsPBC}, we remark that all the conclusions drawn in this Section are independent of the (macroscopic) size $L$ of the system. In fact, even though all the derivation of Eq.~\eqref{eq:equivalence_sigma} is done at finite size, the use of charge fluxes which are well-defined in PBC ensures that the GK integrals of their correlation functions are well-defined and non-vanishing identically, for any $L$.}
\section{Conclusions} \label{sec:conclusions}
To conclude, we believe that we managed to show how the invariance principles of transport coefficients can be employed, within the general Green-Kubo theory of linear response, to demystify some common misconceptions based on the groundless assumption that the microscopic conserved fluxes must be uniquely defined.
We have further deployed the power of invariance principles in the construction of numerical tools for the statistical analysis of the time series of the fluxes, produced via equilibrium molecular dynamics simulations. These tools provide accurate values for the transport coefficients from the relatively short trajectories which are accessible to ab initio MD simulations. We also showed how to design an ab initio heat flux suitable to extended systems in PBC, from which the ab initio thermal conducitivity can be computed without any further system- or phase-specific assumptions. 
Finally, we have applied the gauge invariance principle to the ab initio charge transport in ionic systems. We showed that each ion can be associated to a well-defined integer and time-independent charge, and the exact \textit{ab-initio} electrical conductivity can be obtained by replacing, with these integer charges, the time-dependent Born tensors, which enter the definition of the charge flux and whose calculation is a computationally-expensive and quite abstruse task. In this way, we recover the classical (Faraday's) idea of atomic contributions to charge transport \cite{Resta2021} and provided a theoretically sound definition to the concept of oxidation states in ionic liquid insulators.
We are confident that the results here exposed will be a valid aid for both theorists and practitioners aiming at deeper insights and more efficient implementations about the calculation of transport coefficients from molecular dynamics simulations. 

{
\section*{Data Availability}
The data that support the plots and relevant results within this paper are available on the Materials Cloud platform\cite{talirz2020materials}, DOI:\href{https://doi.org/10.24435/materialscloud:rp-cd}{10.24435/materialscloud:rp-cd}.
}

\section*{Acknowledgements}

We thank Paolo Pegolo for fruitful discussions and a thorough reading of the manuscript. This work was partially funded by the European Union through the \textsc{MaX} Centre of Excellence for Supercomputing applications (Project No. 824143),  by the Italian MIUR/MUR through the PRIN 2017 \emph{FERMAT} grant and by the Swiss National Science Foundation (SNSF), through  Project No. 200021-182057. {FG acknowledges funding from the European Union's Horizon 2020 research and innovation programme under the Marie Sk\l{}odowska-Curie Action IF-EF-ST, grant agreement no. 101018557 (TRANQUIL).}

This is a preprint of an article published in EPJB. The final authenticated version is available online at:\url{http://doi.org/10.1140/epjb/s10051-021-00152-5}.

\section*{Author Contribution Statement}
The authors contributed equally to all parts of this work.

\appendix
\section{Variance on GK and HE formulas} \label{App:A}
\begin{figure}[h!]
    \includegraphics[width=0.95\columnwidth]{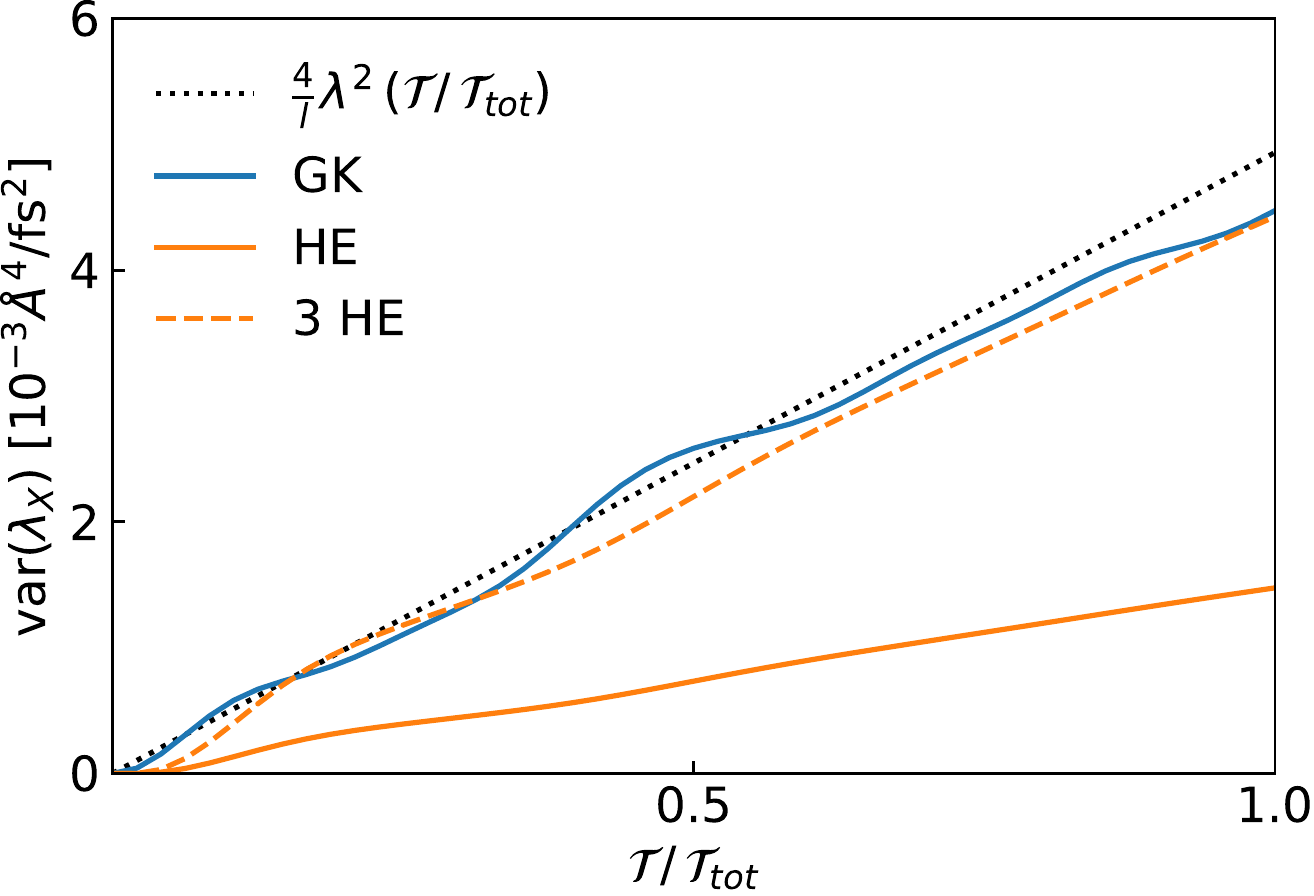}
    \caption{Variance of the estimators of the GK and HE integrals of the charge-flux autocorrelation function, computed from standard block analysis of an ab initio MD trajectory of molten KCl. The variance on the GK integral (blue) is $3$ times the variance affecting the HE formula (orange). The black dotted line displays the theoretical estimate, Eq.~\eqref{eq:varGKHE}.}
    \label{fig:errGKHE}
\end{figure}

In order to evaluate the statistical error affecting the GK or HE expressions of the transport coefficients, Eqs.~\eqref{eq:lambda_X}, we consider it as the expectation of the estimator:
\begin{equation}
    \begin{aligned}
      \stoc{\lambda}_X(\upt) &= \frac{1}{4\pi} \int_{-\infty}^{\infty} \tilde \Theta_X^\upt(\omega) \stoc{S}(\omega) d\omega, \\
      &\approx \frac{1}{2\mathcal{T}_{tot}} \sum_{k=0}^{N-1}\tilde\Theta_X^\upt(\omega_k)S(\omega_k)\stoc\xi_k,
    \end{aligned}
\end{equation}
where $X=GK\text{ or }HE$, $\mathcal{T}_{tot}=N\epsilon$ is the total length of the time series of the flux, $N$ the number of its terms, $\epsilon$ the sampling period, and $\stoc\xi_k$ are the set of independent stochastic variables introduced in Eq.~\eqref{eq:periodogram-distribution}. As the $\stoc\xi_k$ variables are independent and identically distributed, one has:
\begin{equation}
  \begin{split}
    \text{var}\left (\stoc{\lambda}_X \right ) &= \frac{2}{4\mathcal{T}_{tot}^2} \sum_{k=0}^{N-1} \tilde{\Theta}^\upt_{X}(\omega_k)^2 S(\omega_k)^2 \text{var}\left (\stoc\xi \right) \\
    &\approx \frac{1}{2\ell\mathcal{T}_{tot}}\int_{-\infty}^\infty \tilde{\Theta}^\upt_{X}(\omega)^2 S(\omega)^2 \frac{d\omega}{2\pi} \\
    & \approx \frac{S(0)^2}{2\ell\mathcal{T}_{tot}}\int_{-\upt}^\upt {\Theta}^\upt_{X}(t)^2 dt =  \begin{cases} \frac{4}{\ell}\lambda^2 \frac{\upt}{\mathcal{T}_{tot}} & \text{(GK)} \\[3pt] \frac{4}{3\ell}\lambda^2\frac{\upt}{\mathcal{T}_{tot}} & \text{(HE),} \end{cases}
  \end{split}\label{eq:varGKHE}
\end{equation}
where we employed Eq.~\eqref{eq:lambdaSmezzi} and where the factor 2 in the first step accounts for the full correlation between $\stoc{S}(\omega)$ and $\stoc{S}(-\omega)$.
This behaviour is shown in Fig.~\ref{fig:errGKHE}, which displays the theoretical estimate, Eq.~\eqref{eq:varGKHE}, for the variance on the GK integral, as well as the empirical variances for GK (blue) and HE (orange) integrals of the charge flux autocorrelation function, obtained via standard block analysis from the simulation of molten KCl already discussed in Sec.~\ref{sec:theory}. 

Notice that, in order to obtain the variance on the mean value $\langle \stoc{\lambda}_X \rangle$, the variance of the process, which is independent of the number $B$ of trajectories (or blocks) of length $\mathcal{T}_{tot}$ employed, must be further divided by $B$ \cite{Jones2012}.

\FloatBarrier

\end{document}